\documentclass[twocolumn,twoside,slac_two]{revtex4}
\usepackage{graphicx}
\usepackage{fancyhdr}
\def\simleq{\; \raise0.3ex\hbox{$<$\kern-0.75em \raise-1.1ex\hbox{$\sim$}}\; }
\def\simgeq{\; \raise0.3ex\hbox{$>$\kern-0.75em \raise-1.1ex\hbox{$\sim$}}\; }

\newcommand{\TeV}{{\rm TeV}}

\newcommand{\km}{{\rm km}}

\newcommand{\s}{{\rm s}}

\begin{document}

\title{Cosmic ray models compared to Fermi-LAT  positron and electron separate spectra}
\author{D. Grasso, D. Gaggero for the Fermi-LAT collaboration}
\affiliation{INFN Pisa,  Largo B. Pontecorvo, 3, I-56127 Pisa, Italy}
%


\begin{abstract}
In this symposium the Fermi-LAT collaboration released the results of its measurement of the $e^-$ and $e^+$ separate spectra, obtained using the Earth magnetic field. Those results confirm PAMELA finding of an increasing positron fraction above 10 GeV and allow, for the first time independently on the electron spectrum, to reject cosmic ray spallation onto the ISM as the dominant positron production mechanism above that energy.  We show, for two different propagation setups, that double component models which were developed to provide a consistent interpretation of PAMELA positron fraction and Fermi-LAT $e^+ + e^-$ data,  naturally reproduce also the Fermi-LAT $e^-$ and $e^+$ separate spectra.
\end{abstract}

\maketitle

\thispagestyle{fancy}

\section{Introduction}

One of the most striking recent results in cosmic rays (CR) physics was the observation performed by the PAMELA satellite experiment that the positron to electron fraction $e^+/ (e^- + e^+)$  rises with energy from 10 up to  100 GeV at least \cite{Adriani:2008zr}. 

In the standard CR scenario, the bulk of electrons reaching the Earth in the GeV - TeV energy range are originated by Supernova Remnants (SNRs) and only a subdominant component of secondary positrons and electrons comes from the interaction of CR nuclei with the interstellar medium (ISM). 
Since the residence time of primary nuclei in the Galaxy decreases with energy, in that scenario a growing positron fraction could be justified only by a very steep primary electron spectrum. This possibility, however, was ruled out by the Fermi-LAT collaboration which found the $e^- + e^+$ spectrum in the 7 GeV - 1 TeV energy range to be compatible with a power-law with index $\gamma(e^\pm) = - 3.08 \pm 0.05 $  \cite{Abdo:2009zk,Ackermann:2010ij}, which is significantly harder than what estimated on the basis of previous measurements.  
Although excluded by the PAMELA collaboration, in principle, the rising positron fraction could be the consequence of misidentified CR protons (which are $10^4 \div 10^5$ times more abundant than positrons).  Therefore, an independent confirmation of PAMELA results was called for. 

The awaited confirmation arrived just during this symposium as the Fermi-LAT collaboration released the preliminary results of the measurements of the absolute $e^+$ and $e^-$
spectra, and of their fraction, between 20 and ~120 GeV performed using the EarthÕs magnetic field (W. Mitthumsiri et al. plenary talk). More recently, the collaboration published the final results of those measurements which were extended up to 200 GeV \cite{:2011rq}. The sum of the $e^+$ + $e^-$ separate spectra was found to agree with the $e^+ + e^-$ spectrum previously measured by Fermi-LAT.
A steady rising of the positron fraction was observed by this experiment up to that energy in agreement with that found by PAMELA.  In the same energy range, the $e^-$ spectrum was fitted with a power-law with index $\gamma(e^-) = - 3.19 \pm 0.07$ which is in agreement with what recently measured by PAMELA between 1 and  625 GeV \cite{Adriani:2011xv}.  Most importantly, Fermi-LAT measured, for the first time, the $e^+$ spectrum in the 20 - 200 GeV energy interval and showed it is fitted by a power-law with index  $\gamma(e^+) =  - 2.77  \pm 0.14$.  

Remarkably, the $e^+$ spectral index measured by Fermi-LAT is almost coincident with that of CR protons which, in the standard scenario, are the main positron primaries. 
Since Feynman scaling implies that secondary positrons are produced with the same spectral index of protons, no room is then left to the steepening that propagation and energy looses should unavoidably produce in the propagated $e^+$ spectrum.  From this simple argument it is already evident that the standard $e^+$ production scenario must be incomplete. 
In the following we will explicitly show this is the case for two different propagation setups. We will then show to which extent and under which conditions the alternative scenario, which invokes the presence of a new primary positron and electron component, consistently describes all Fermi-LAT and PAMELA electron and positron data.   

\section{Single vs double component models}\label{sec:models}

In \cite{Abdo:2009zk,Grasso:2009ma,Ackermann:2010ij} the Fermi-LAT collaboration discussed two scenarios to describe the $e^+ + e^-$ spectrum it measured.  Schematically, we can define them as single and double component scenarios. 

\begin{figure*}[t]
\centering
\includegraphics[width=85mm]{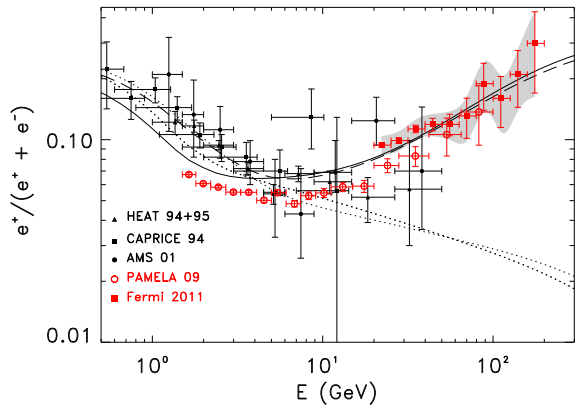}
\caption{The positron fraction computed with {\tt DRAGON} or {\tt GALPROP} is compared to experimental data. The shaded region represents the sum in quadrature of statistical and systematical Fermi-LAT errors.
Upper/lower dotted lines represent the predictions of the standard single component scenario for strong/moderate reacceleration propagation setups described in Sec.\ref{sec:models}/\ref{sec:kra_model}. Dashed/solid lines have been derived for double component models using the same propagation setups respectively. 
All models assume solar modulation with potential $\Phi = 550~$ MV.  } 
\label{fig:pos_fraction}
\end{figure*}

The former scenario assumes one single class of sources, spatially distributed like Galactic SNR's, injecting $e^-$ in the ISM with a, {\it a priori} unknown, broken power-law spectrum. 
In those papers propagation was modeled using the {\tt GALPROP} \cite{Strong:1998pw} numerical diffusion package and solar modulation was then taken into account in order to reproduce low energy data. This was done in the charge independent force field approximation by fixing the modulation potential $\Phi$ against proton data taken in the same solar phase.   
 
Both models considered in \cite{Ackermann:2010ij} assume a cylindrical diffusive halo with half-thikness of 4 kpc, a diffusion coefficient scaling with rigidity like $\rho^{1/3}$ (Kolmogorov like diffusion) and relatively strong reacceleration  (the Alfv\'en velocity was taken $v_A = 30~\km \s^{-1}$). Under those conditions, {\tt GALPROP} provides an excellent description of most CR nuclei measurements \cite{Strong:2007nh}. It should be taken in mind, however, that the observed antiproton spectrum is not well fitted under those conditions while other choices of the propagation parameters were shown to consistently reproduce both nuclear and antiprotons data (see Sec. \ref{sec:kra_model}).  
 
\begin{figure*}[t]
\centering
\includegraphics[width=85mm]{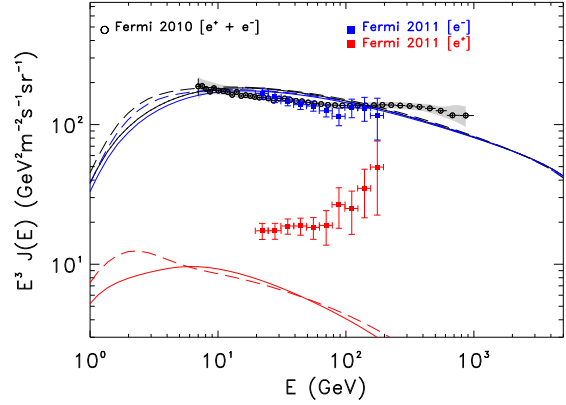}
\caption{The $e^+ + e^-$ (black lines), the $e^+$ (red lines) and the $e^-$ (blue lines) spectra computed for the single component scenario are compared with Fermi-LAT data in \cite{:2011rq} .  
Dashed/solid lines have been derived using the strong/moderate reacceleration setups described in Sec.\ref{sec:models}/\ref{sec:kra_model}. 
 Statistical and systematic errors are summed in quadrature. All lines accounts for solar modulation with $\Phi = 550~$ MV. }
\label{fig:single_model}
\end{figure*}

The single component reference model in \cite{Ackermann:2010ij} is characterized by an $e^-$ injection spectral index $\gamma_0(e^-) = -1.6/-2.5$ below/above 4 GeV. 
That spectral break is required to reproduce low energy AMS-01 $e^-$ data  \cite{ams1} as well as the spectrum of the synchrotron emission of the Galaxy below 1 Ghz \cite{Strong:2011wd}.
This model differs from similar pre-Fermi {\tt GALPROP} models just for the harder spectrum it adopts above 4 GeV as required to track the Fermi-LAT $e^+ + e^-$ hard spectrum.
An high energy cutoff in the source spectrum was also introduced in order not to overshoot H.E.S.S. $e^+ + e^-$ data \cite{Aharonian:2009ah} in the TeV region. 
In spite of such tuning this model does not allow a very satisfactory fit of the $e^+ + e^-$ observed spectrum. Furthermore, since it assumes only $e^+$ secondary production, it cannot explain the positron fraction rise observed by PAMELA and now confirmed by Fermi-LAT  (see dotted lines in Fig. \ref{fig:pos_fraction} ). 
In Fig. \ref{fig:single_model}  \footnote{The models represented in all figures of this contribution have been computed with {\tt DRAGON} \cite{dragon}.  
We verified that {\tt DRAGON} gives the same {\tt GALPROP v54} \cite{Vladimirov:2010aq}  results under equal conditions.} we show that model also fails to describe the $e^+$ measured by Fermi-LAT. This is the case also for the moderate reacceleration propagation setup that will be discussed in the next section. 

The double component scenario is characterized by the presence of a new term in the electron and positron source spectrum with the form 
\begin{equation}
J_{\rm extra}(e^\pm) \propto E^{~\gamma_0({e^\pm})}~\exp(- E/E_{\rm cut})~.
\label{eq:extra_comp}
\end{equation}
For $\gamma_0({e^\pm})\simeq - 1.5$ and  $E_{\rm cut} \simeq 1~\TeV$  this term has been shown to account not only for the positron fraction anomaly observed by PAMELA 
\cite{Hooper:2008kg} but also an excellent fit of the $e^+ + e^-$ spectrum measured by Fermi-LAT and H.E.S.S.  \cite{Grasso:2009ma,Ackermann:2010ij}. 
Several hypothesis have been risen for its origin including $e^\pm$ acceleration in pulsar wind nebulae, dark matter annihilation, secondary $e^\pm$ production in SNRs 
(see \cite{Grasso:2009ma} and Ref.s therein).
Although the spatial distribution of the extra component source term generally depends on which of those scenarios is adopted, this has no consequences below few hundred GeV since at those energies the $e^\pm$ propagation length is comparable to the Galaxy size so that spatial features in the source term are averaged-out.   
For this reason the following considerations apply both to astrophysical and to dark matter double component models.   

\begin{figure*}[t]
\centering
\includegraphics[width=85mm]{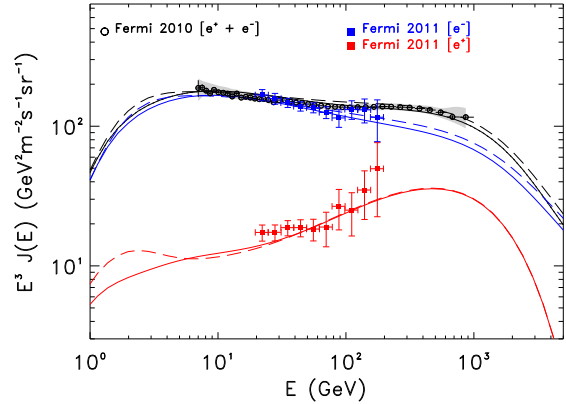}
\caption{The predictions of the double component models discussed in Sec.s \ref{sec:models} and \ref{sec:kra_model} are compared with Fermi-LAT data in \cite{:2011rq}.  The modulation potential and the line notation are the same of Fig. \ref{fig:single_model}. }
\label{fig:double_model}
\end{figure*}

The double component model considered in \cite{Ackermann:2010ij} adopts $\gamma_0(e^-) = -1.6/-2.7$ below/above 4 GeV for the standard electron component, $\gamma_0({e^\pm}) = - 1.5$ and $E_{\rm cut} = 1.4~\TeV$ for the electron and positron extra component.
In Fig.s \ref{fig:pos_fraction} and \ref{fig:double_model}  we compare the predictions of that model also with the new Fermi-LAT data \cite{:2011rq}.
The reader can see from those figures as this model correctly reproduces not only the $e^+ + e^-$ spectrum measured by Fermi-LAT but also its new $e^-$ spectrum.  
The model also matches most Fermi-LAT $e^+$ data points and, most importantly, it reproduces their slope. 
The new Fermi-LAT data, therefore, confirm the presence of an $e^\pm$ spectral component in the form given in Eq.\ref{eq:extra_comp}.

\section{The case of moderate reacceleration}\label{sec:kra_model}

In this section we test to which extent the choice of the propagation setup may affect our previous results. 
Our attempt is also motivated by the well known discrepancy between the prediction of strong reaccelation models and the positron fraction measured by PAMELA below 10 GeV.  Such {\it low energy PAMELA anomaly} is commonly ascribed to solar modulation behaving in a more involved way than in the force field approximation. Charge dependent effects correlated to the solar magnetic field polarity are indeed theoretically expected as a consequence of the complex spatial and temporal structure of the heliosphere (see \cite{Potgieter}  for a recent review).Typically, however, those effects are relevant only in the sub-GeV region and a satisfactory description of all available electron and positron data in those terms has not been yet achieved. 

Here we suggest that at least part of the low energy positron fraction anomaly could be related to other issues encountered by strong reacceleration propagation setups:  
1) as we mentioned, those models do not provide a satisfactory description of the antiproton spectrum measured by PAMELA and other experiments \cite{Strong:2007nh,DiBernardo:2009ku};  2) they reproduce the observed proton spectrum only by introducing a break {\it ad hoc} in the source spectrum;  3)  strong reacceleration rise similar problems with the electron spectrum in the GeV region \cite{DiBernardo:2010is} ; 4) they are incompatible with the observed spectrum of the synchrotron emission of the Galaxy \cite{Strong:2011wd}. 
Although none of these warnings might exclude strong reacceleration by itself, taken together they seriously justify to consider alternative possibilities.

For all these reasons we consider here also a propagation setup characterized by moderate reacceleration ($v_A = 15~\km \s^{-1}$) and a diffusion coefficient with a Kraichnan like dependence on rigidity:  $D(\rho) \propto \rho^{1/2}$. More specifically we adopt here the KRA model proposed in \cite{DiBernardo:2010is}.
That model was shown to correctly reproduce the B/C and other nuclear CR data if the particle speed dependence of the diffusion coefficient is suitably tuned (as could be justified by MHD wave dissipation,  see \cite{Strong:2007nh} and ref.s therein).  Differently from strong reacceleration setups, it reproduces the antiproton spectra measured by PAMELA  \cite{DiBernardo:2009ku} and, what is most relevant here, it requires no break in the proton source spectrum at least up to few hundred GeV. 
In fact, we verified that this is crucial to correctly match the PAMELA positron fraction below 10 GeV. 
Under those conditions, and assuming the presence of a $e^\pm$ extra-component as that adopted in the previous section, the $e^+ + e^-$ spectrum measured by Fermi-LAT, H.E.S.S. and AMS-01 was also accurately reproduced.  Respect to Ref. \cite{DiBernardo:2010is} here we only change the low energy $e^-$ source spectral index by taking  $\gamma_0(e^-) = -1.6/-2.65$ below/above 4 GeV rather that $-2/2.65$.  
In Fig. \ref{fig:pos_fraction} the reader can see as this arrangement allows not only a good fit of the $e^+$,  $e^-$ and $e^+/(e^+ + e^-)$ Fermi-LAT data between 20 and 200  GeV but also to improve considerably the description of the PAMELA positron fraction data below 20 GeV. The small residual offset may either be due to solar modulation or to a experimental systematic (the reader may have noted the small offset between the $e^+$ fraction measured by PAMELA and Fermi-LAT).  

We verified that PAMELA $e^-$ data \cite{Adriani:2011xv} are also consistently reproduced by the model considered in this section, with almost the same modulation potential, if their normalization is rescaled so to make them to coincide with Fermi-LAT $e^-$ data at 20 GeV. 

\section{Conclusions}

In this symposium the Fermi-LAT collaboration presented its measurements of the positrons fraction as well as of the $e^+$ and $e^-$ absolute spectra in the 20 - 200 GeV energy  interval.  We showed that the $e^+$ spectrum measured by Fermi-LAT is in contrast with the standard scenario in which $e^+$ are originated only by the spallation of the nuclear component of CR onto the ISM. This evidence is even stronger than that provided by the positron fraction anomaly found by PAMELA, and confirmed by Fermi-LAT, since it is independent on the $e^-$ spectrum reducing the involved systematics. We then tested the alternative scenario which invoke the presence of an additional electron and $e^+$ primary spectral component, against the same data. 
We showed, for two different propagation setups, that models which were proposed in this framework to consistently interpret PAMELA positron and the Fermi-LAT $e^+ + e^-$ spectrum naturally reproduce the  $e^+$ and $e^-$ separate  spectra. By measuring the $e^+$ and $e^-$ spectra above 200 GeV, AMS-02 will be necessary to have a clue on the nature of the extra component discovered by PAMELA and Fermi-LAT.  

\section*{Acknowledgments}

The Fermi LAT Collaboration acknowledges generous ongoing support from a number of agencies and institutes that have supported both the development and the operation of the LAT as well as scientific data analysis. These include the National Aeronautics and Space Administration and the Department of Energy in the United States, the Commissariat ˆ l'Energie Atomique and the Centre National de la Recherche Scientifique / Institut National de Physique NuclŽaire et de Physique des Particules in France, the Agenzia Spaziale Italiana and the Istituto Nazionale di Fisica Nucleare in Italy, the Ministry of Education, Culture, Sports, Science and Technology (MEXT), High Energy Accelerator Research Organization (KEK) and Japan Aerospace Exploration Agency (JAXA) in Japan, and the K. A. Wallenberg Foundation, the Swedish Research Council and the Swedish National Space Board in Sweden.



\begin{thebibliography}{9}

\bibitem{Adriani:2008zr}
  O.~Adriani {\it et al.}  [PAMELA Collaboration],
  Nature {\bf 458} (2009) 607
  [arXiv:0810.4995 [astro-ph]].


\bibitem{Abdo:2009zk}
  A.~A.~Abdo {\it et al.}  [The Fermi LAT Collaboration],
  Phys.\ Rev.\ Lett.\  {\bf 102}, 181101 (2009)
  [arXiv:0905.0025 [astro-ph.HE]].

\bibitem{Ackermann:2010ij}
  M.~Ackermann {\it et al.}  [Fermi LAT Collaboration],
  Phys.\ Rev.\  D {\bf 82} (2010) 092004
  [arXiv:1008.3999 [astro-ph.HE]].

\bibitem{:2011rq}
  M.~Ackermann {\it et al.}   {\it et al.} [ The Fermi LAT Collaboration ],
   [arXiv:1109.0521 [astro-ph.HE]].

\bibitem{Adriani:2011xv}
  O.~Adriani {\it et al.}  [PAMELA Collaboration],
  Phys.\ Rev.\ Lett.\  {\bf 106} (2011) 201101
  [arXiv:1103.2880 [astro-ph.HE]].

\bibitem{Hooper:2008kg}
  D.~Hooper, P.~Blasi, P.~D.~Serpico,
  JCAP {\bf 0901 } (2009)  025.
  [arXiv:0810.1527 [astro-ph]].

\bibitem{Grasso:2009ma}
  D.~Grasso {\it et al.}  [FERMI-LAT Collaboration],
  Astropart.\ Phys.\  {\bf 32} (2009) 140
  [arXiv:0905.0636 [astro-ph.HE]].

\bibitem{Strong:1998pw}
  A.~W.~Strong and I.~V.~Moskalenko,
  Astrophys.\ J.\  {\bf 509} (1998) 212
  [arXiv:astro-ph/9807150].

\bibitem{Strong:2007nh}
  A.~W.~Strong, I.~V.~Moskalenko and V.~S.~Ptuskin,
  Ann.\ Rev.\ Nucl.\ Part.\ Sci.\  {\bf 57} (2007) 285
  [arXiv:astro-ph/0701517].

\bibitem{ams1} M.~Aguilar {\it et al.}, Phys.\ Reports {\bf 366} (2002) 331.

\bibitem{Vladimirov:2010aq}
  A.~E.~Vladimirov {\it et al.},
  Comput.\ Phys.\ Commun.\  {\bf 182} (2011) 1156
  [arXiv:1008.3642 [astro-ph.HE]].

\bibitem{dragon}  {\tt DRAGON} web page: http://www.desy.de/~maccione/DRAGON/

\bibitem{Aharonian:2009ah}
  F.~Aharonian {\it et al.}  [H.E.S.S. Collaboration],
  Astron.\ Astrophys.\  {\bf 508} (2009) 561
  [arXiv:0905.0105 [astro-ph.HE]].

\bibitem{Strong:2011wd}
  A.~W.~Strong, E.~Orlando and T.~R.~Jaffe,
 A\&A\  {\bf 534}  (2011) A54 
  [arXiv:1108.4822 [astro-ph.HE]].
  
\bibitem{Potgieter} M.~Potgieter, S.~Ferreira and D.T.~Strauss,  Proceed. of the 12th ICATPP conference, Como, Oct. 2010  pag. 441   

\bibitem{DiBernardo:2010is}
  G.~Di Bernardo  {\it et al.}, 
  Astropart.\ Phys.\  {\bf 34} (2011) 528
  [arXiv:1010.0174 [astro-ph.HE]].
  
\bibitem{DiBernardo:2009ku}
  G.~Di Bernardo, C.~Evoli, D.~Gaggero, D.~Grasso, L.~Maccione,
  Astropart.\ Phys.\  {\bf 34 } (2010)  274-283.
  [arXiv:0909.4548 [astro-ph.HE]].


\end{thebibliography}
\end{document}